# Application of machine learning in grain-related clustering of Laue spots in a polycrystalline energy dispersive Laue pattern

Amir Tosson*, Mohammad Shokr,
Mahmoud Al Humaidi, Eduard Mikayelyan,
Christian Gutt and Ulrich Pietsch

Department of Physics,
The University of Siegen,
Walter-Flex-Str. 3, 57072,
Siegen, Germany
Email: amir.tosson@uni-siegen.de
Email: mohammadshokr@hotmail.com
Email: mahmoud.alhumaidi@hotmail.com
Email: edmikayelyan@gmail.com
Email: christian.gutt@uni-siegen.de
Email: pietsch@physik.uni-siegen.de
*Corresponding author

**Abstract:** We address the identification of grain-corresponding Laue reflections in energy dispersive Laue diffraction (EDLD) experiments by formulating it as a clustering problem solvable through unsupervised machine learning (ML). To achieve reliable and efficient identification of grains in a Laue pattern, we employ a combination of clustering algorithms, namely hierarchical clustering (HC) and K-means. These algorithms allow us to group together similar Laue reflections, revealing the underlying grain structure in the diffraction pattern. Additionally, we utilise the elbow method to determine the optimal number of clusters, ensuring accurate results. To evaluate the performance of our proposed method, we conducted experiments using both simulated and experimental datasets obtained from nickel wires. The simulated datasets were generated to mimic the characteristics of real-world EDLD experiments, while the experimental datasets were obtained from actual measurements.

**Biographical notes:** Amir Tosson is a distinguished professional in the field of Software Engineering, known for his expertise in artificial intelligence and machine learning. With a career spanning over a decade, he has contributed to numerous innovative projects, shaping the future of technology. He holds a Master's and PhD degrees in Applied Physics from the University of Siegen and has authored several influential papers in top-tier journals. He is also an active mentor, guiding technologists in their careers. Beyond his professional life, he is passionate about community service and regularly participates in initiatives aimed at bridging the digital divide.

Mohammad Shokr earned his PhD in Physics with a focus on photonics, detector physics, and material characterisation from the University of Siegen in 2019. His dissertation, titled "From pnCCD to pnCCD + CsI(Tl) Scintillator: Characterizations and Applications", reflects his expertise. He conducted postdoctoral research at the University of Siegen from 2019 to 2021. Since 2022, he has been a detector development scientist at EuXFEL in Hamburg, where he continues to advance the field with his innovative work.

Mahmoud Al Humaidi earned his PhD in Physics from the University of Siegen in 2021. Following his doctoral studies, he worked as a postdoctoral researcher at the Karlsruhe Institute of Technology from 2021 to 2022. His research focuses on advanced topics in physics, contributing to significant developments in his field.

Eduard Mikayelyan earned his PhD in Physics from the University of Siegen. He significantly contributed to the design and development of a cell for operando analysis of organic semiconductors using X-ray analysis. This innovation earned a beamtime award from the European Synchrotron Radiation Facility in 2012 for a joint project on structure- properties correlation in novel semiconductors for organic electronics.

Christian Gutt is a Professor at the University of Siegen, known for his expertise in experimental physics. His research focuses on the study of condensed matter physics, particularly using X-ray and synchrotron radiation techniques. He has made significant contributions to understanding the structural dynamics of complex materials.

Ulrich Pietsch is a Distinguished Professor at the University of Siegen. He has made significant contributions to the field of materials science and solid state, particularly in the study of organic semiconductors and X-ray analysis techniques. His research has been instrumental in advancing the understanding of structure-properties correlations in novel semiconductors, leading to several high-impact publications and international collaborations.

## 1 Introduction

X-ray diffraction techniques are powerful tools for elemental, chemical, and structural analysis of polycrystalline materials (Kirkwood et al., 2017). Taking advantage of white synchrotron radiation from high brilliance storage ring sources in combination with fast pixelated area detectors, it is nowadays possible to map a large volume of reciprocal space of the sample within a single exposure diffraction pattern (Rakowski et al., 2017; Abboud et al., 2017; Brinkmann, 2007). One such technique is energy dispersive Laue diffraction (EDLD) (Send et al., 2009; Ordavo et al., 2011), which is a non-destructive tool to investigate the structural characteristics of monocrystalline and polycrystalline materials which requires only little efforts for sample alignment and orientation. EDLD is based on Bragg's law, whereby both the angular positions and the diffracting energies of the Laue spots can be analysed without additional information, allowing for a quick identification of the crystal structure and the respective lattice parameters. Due to its low cost, rapid elemental analysis, and quasi one-shot mode of operation, EDLD has emerged as an attractive and dependable tool in various fields of the material sciences (Send et al., 2016). This technique proves particularly valuable in scenarios where quantifying the composition of complex and locally inhomogeneous materials is essential, such as when examining materials influenced by the manufacturing process history. This technique proves particularly valuable in scenarios where quantifying the composition of complex and locally inhomogeneous materials is essential, such as when examining materials influenced by the manufacturing process history.

In this study, we introduce a new method with practical implications and wide-ranging applications in understanding and using grain structures. By improving how we precisely and efficiently gather information about the structure of grains in EDLD, our research directly adds to our fundamental knowledge of material properties and behaviour. Understanding grain structures is crucial in fields like materials science, metallurgy, and geophysics.

Moreover, the method we propose is not limited to crystallography – it can be useful in any field where identifying and characterising distinct entities within complex patterns is important. For example, in image analysis, this approach could help identify and categorise objects in detailed visual data. It might also be valuable in biology for analysing cellular structures or in social sciences for recognising patterns in diverse datasets. In summary, our method is a flexible tool that boosts our ability to get meaningful information from intricate patterns, contributing to progress in various scientific and technological areas.

## *1.1 Experimental setup*

Since the EDLD is an angle- and energy-dispersive tool, the sample alignment is straightforward and does not require sample rotation, beam alignment with respect to the sample, or other preparational steps. Figure 1 shows the typical experimental setup of EDLD experiments. White X-ray radiation is usually provided by a storage rings or conventional X-ray tubes and ranges typically from 5 keV up to about 140 keV photon energy. A collimating system is used to tailor the beam size and an absorber system enables to tune both X-ray spectrum and X-ray flux, with the aim to run an energy-dispersive 2D detection system operating in the so called 'single photon counting mode' as the pnCCD system, for example (Shokr et al., 2017). The experiment can be performed either in transmission or reflection geometry where the sample is always located close to the front of the detector. Both the sample and the detector may be equipped by high resolution motor stages with steps sizes down to 1µm, enabling to change the scattering geometry in steps of microns. During the exposure time, the incident beam is elastically scattered by the sample and the diffraction spots are collected on the detector. Due to the large bandwidth of the white X-ray beam, a characteristic multi-reflections Laue pattern is formed. This pattern may include X-ray fluorescence signals from the chemical elements present in the sample and from the equipment. Importantly, it contains the intense reflections, often called Laue spots, originating from scattering at crystal planes which fulfil the Bragg condition. This EDLD setup is ideally suited for investigations of the micro-structure of polycrystalline materials, such as metal and ceramics. These materials are typically formed from combinations of many single-crystalline grains. Each grain exhibits a certain orientation with respect to the incident beam and an external reference system of coordinates.

**Figure 1** The EDLD experimental setup (see online version for colours)

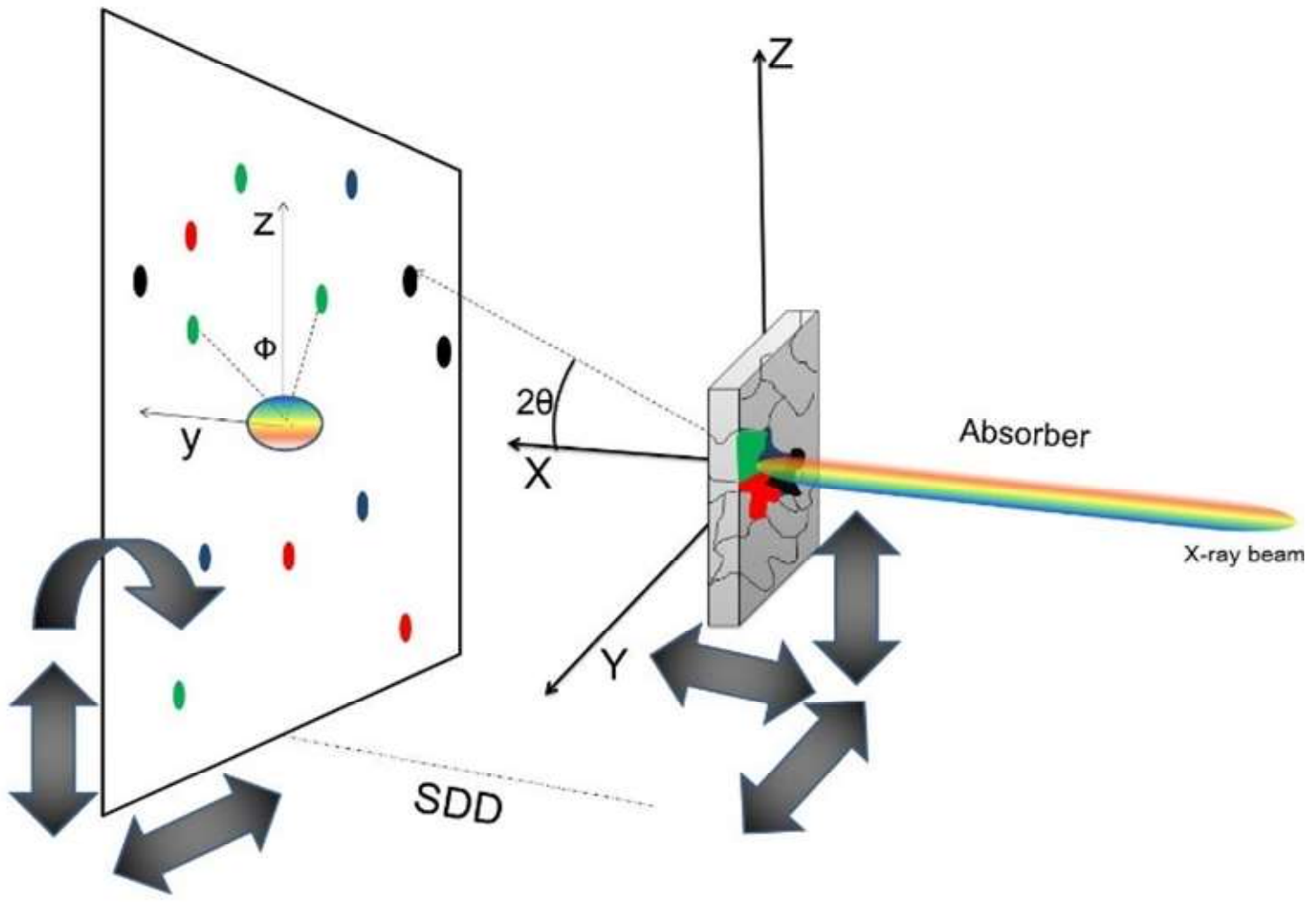

Notes: X, Y and Z represent the laboratory system and (y, z) are the detector coordinates. The incident beam is parallel to X-axis. Each Laue spot has a specific position at the detector active area (y_i, z_i). Φ is the angle between two diffracted spots and SDD denotes the distance between the sample and the detector. In some experiments, the sample could be moved in directions X, Y and Z with respect to the chosen reference position. The detector could be moved in three directions.

## *1.2 The bottleneck of large data volumes*

The 5D characteristic (x-position, y-position, energy, intensity, and time) of EDLS carries a wealth of information, which, however, is not easily accessible due to the huge computational workload needed for analysing and indexing diffraction patterns with a substantial overlap of diffraction peaks generated from differently orientated grains. This in turn requires highly efficient computational techniques which group overlapping spots and identify their corresponding grain. This problem is commonly known as the grain-related clustering of Laue spots. As conventional analysis methods are often slow and not fully reliable, there have been attempts to develop more effective and systematic techniques that can perform the grouping (Chung and Ice, 1998; Tosson et al., 2019; Gupta and Agnew, 2009). In our previous work (Tosson et al., 2019), we introduced parallel programming and high-performance computing as a powerful solution to this problem. However, this approach requires significant computational resources and infrastructure. Therefore, it is less attractive to the scientific communities. In addition, it is based on the availability of a minimum of three collected reflections per grain to allow for the recognition of an individual grain. Here, we demonstrate a new machine-learning approach that does not require special IT resources and can execute the task with high efficiency, speed, and accuracy. Moreover, it has no limitation regarding the number of collected reflections per grain.

 The novelty of applying machine learning (ML) in grain-related clustering of Laue spots within a polycrystalline energy dispersive Laue pattern lies in its transformative approach to crystallographic analysis. Unlike traditional methods, this innovative approach harnesses the power of ML algorithms to autonomously identify and categorise Laue spots associated with individual grains. By leveraging advanced computational techniques, the system is capable of discerning complex patterns and relationships within the energy dispersive Laue pattern, providing a more accurate and efficient means of characterising the polycrystalline structure. This not only enhances the precision of grain-related clustering but also significantly reduces the time and effort required for manual analysis, marking a substantial leap forward in the field of crystallography.

## *1.3 State-of-the-art*

Extensive efforts have been dedicated to enhancing the efficiency and accuracy of clustering Laue spots using a variety of techniques. Among these methods, those referred to as (Chung and Ice, 1998; Tosson et al., 2019) employ a conventional computing approach based on trial and error. These techniques systematically explore all possible permutations and subsequently select the most optimal one according to a predefined criterion. However, when compared to our novel approach, these methods exhibit certain limitations, particularly concerning the number of reflections that can be gathered and the resulting delays in execution time. In contrast, the alternative approach introduced in Gupta and Agnew (2009) entails a comparison between experimental diffraction patterns and simulated templates corresponding to crystals with predetermined orientations. It is crucial to acknowledge that, when compared to our new approach, this method encounters limitations in EDLD analysis due to the potential unavailability of information about the grain’s orientation.

Aiming to address the limitations of the traditional computing approach in synchrotron data analysis, the adoption of AI-based techniques is quickly gaining momentum (McClure et al., 2020; Dingel et al., 2021; Shi et al., 2022). For example, a deep-learning approach to process scattering data at synchrotron facilities on-the-fly has been introduced in Starostin et al. (2022) and Wang et al. (2016). It provides automated solutions that enable rapid data processing, overcoming the challenges posed by high acquisition rates and time-consuming data analysis. Another approach, the so-called BraggNN, was proposed in Liu et al. (2022). It relies on deep learning techniques to rapidly determine Bragg-peak positions in comparison to traditional pseudo-Voigt peak fitting.

The distinctive feature of the proposed approach lies in its capacity to extract granular information from the Laue pattern, irrespective of its intricacy and occupancy. This innovative method goes beyond conventional techniques by enabling the extraction of detailed information at the grain level, regardless of the complexity and occupancy factors associated with the Laue pattern.

## 2 Grain-related clustering of Laue spots

The primary objective of this work is to develop a novel technique that can accurately solve the problem of clustering Laue spots in polycrystalline ED Laue patterns, while also providing real-time output, high efficiency for big data, and low consumption of computing resources. To achieve this goal, the problem is treated as a traditional clustering problem, where unsupervised learning algorithms can be employed. In this approach, each cluster corresponds to Laue spots that are connected to a specific single grain, with the number of clusters ($k$) indicating the number of grains present in the sample. However, extracting the necessary similarity feature for accurate clustering is not directly possible from the raw data, which only provides information about the reciprocal space.

To address this challenge, we propose a crystal-related step that transforms the laboratory coordinates of each spot of the pattern by a rotation matrix to the coordinates of a reference grain. This process assigns a label called the 'orientation stamp' to each collected spot, which measures the orientation angle of each grain relative to a chosen reference grain in the 3D-space. It should be noted that the crystallographic planes of every grain in the sample are inclined by the same orientation 3D-angle relative to a reference coordinate system. This reference system can be established by selecting any arbitrary grain, later denoted as *reference grain*. Using this approach, all Laue spots that show the same orientation angle with respect to the reference grain are attributed to the same grain. Indeed, this concept of using the orientation angle as a similarity feature can be employed to conduct the clustering process. Figure 2 visualises this concept by displaying two grains of the same material with identical crystal structures, with '*grain 2*' being rotated with respect to the reference grain '*grain 1*'. Two identical planes $(hkl)_1$ and $(hkl)_2$ are displayed in both grains. The vectors $n_1$ and $n_2$ are normal to the planes in '*grain 1*', while $n_3$ and $n_4$ belong to '*grain 2*'. From this perspective, all normal vectors in '*grain 2*' are rotated with the same 3D-angle to their identical norms (twins) in '*grain 1*'. In other words, '*grain 2*' can be considered as a representation of '*grain 1*' in a different coordinate system. The orientation angles between these two systems (of both grains) are connected via:

$$\epsilon_2 = \alpha = \beta$$

$$\alpha = \cos^{-1}\left(\frac{\overrightarrow{n_2} \cdot \overrightarrow{n_4}}{\left\|\overrightarrow{n_2}\right\|\left\|\overrightarrow{n_4}\right\|}\right)$$

$$\beta = \cos^{-1}\left(\frac{\overrightarrow{n_1} \cdot \overrightarrow{n_2}}{\left\|\overrightarrow{n_1}\right\|\left\|\overrightarrow{n_3}\right\|}\right)$$

where $\epsilon_2$ is *the orientation stamp* for '*grain 2*'. It is a fingerprint of each grain within the sample and represents an orientation angle. Each individual grain has a unique orientation stamp which is used as a similarity feature. The n⃗n.n⃗m is the dot product of the $n^{th}$ and $m^{th}$ norms and can be calculated as follows:

$$\overrightarrow{n_m} \cdot \overrightarrow{n_n} = q_{nx}q_{mx} + q_{ny}q_{my} + q_{nz}q_{mz}$$

The magnitude of the nth norm is calculated as follows:

$$\left\|\overrightarrow{n_{3m}}\right\| = \sqrt{q_{nx}^2 + q_{ny}^2 + q_{nz}^2}$$

where $q_x$, $q_y$, and $q_z$ are the $q$-vectors of the reflection and can be determined directly from the data.

**Figure 2** Schematics of two differently orientated grains in the sample with corresponding lattice planes and normal vectors $\overrightarrow{n_l}$ (see online version for colours)

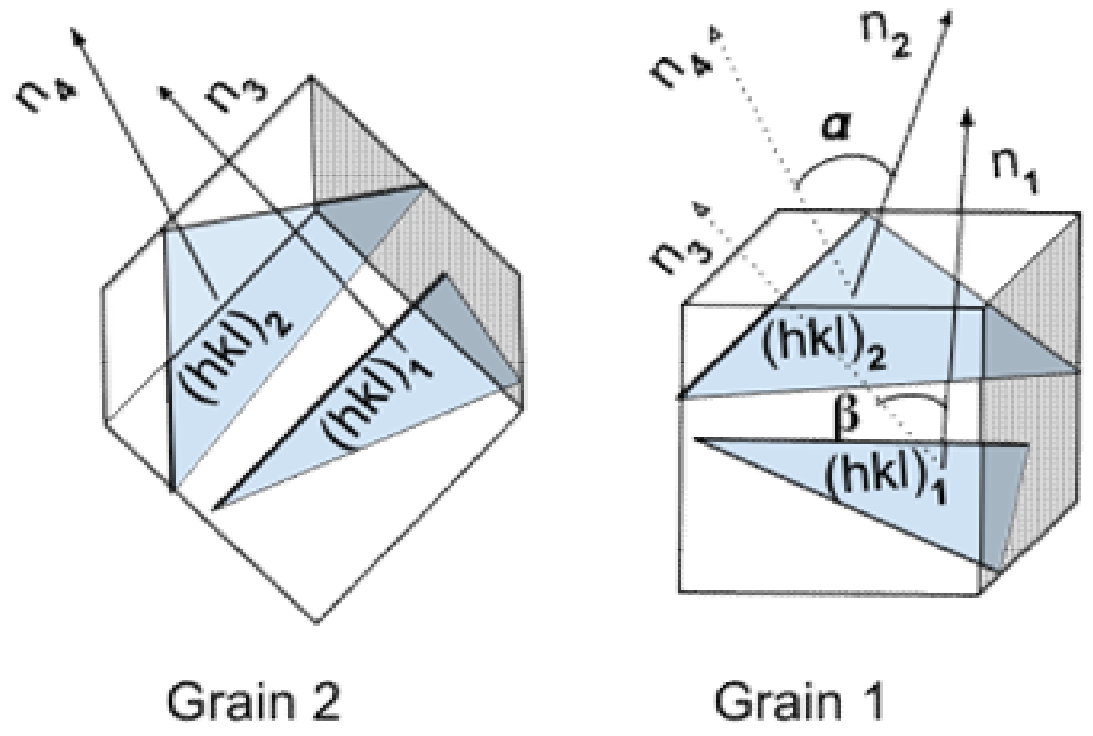

## 3 Data preparation steps

Technically, our approach involves two steps of preparation. Firstly, we must select a reference grain that serves as the basis for comparing the relative orientations of all other grains. Secondly, we need to assign a 3D orientation stamp ($\epsilon_n$) to each Laue spot, where $\epsilon_n$ comprises the three components $\epsilon_x$, $\epsilon_y$, and $\epsilon_z$. Then, the Laue spots are expressed as orientation angles rather than their spatial coordinates. This novel data representation is referred to as the '*AT-map*'.

### *3.1 Reference grain*

Due to the stochastic nature of the collected reflections, a valid reference grain should ideally contain all possible reflections for a given experimental setup. However, in practice, it is challenging to cover a sufficiently large solid angle with the detector, given the limited size of its active area. To address this issue, one possible solution is to simulate the Laue spots from a reference grain as a prepossessing step. To achieve this, a specialised algorithm has been developed, utilising inverse modelling through projection engineering and vector algebra. The process begins by generating an extensive list of all potential reflections that can be produced by a particular crystal structure under specific experimental parameters (such as detector position relative to the incident beam, photon energy range, and Miller indices range). This is accomplished by constructing a reciprocal space map for a bulk crystal and performing 3D rotations on it, if needed. The pseudo-code for the grain simulator is presented in Algorithm 1. For illustration, we created a simulated dataset consisting of two grains of gallium arsenide (GaAs) with Miller indices ranging from −3 to 3. One grain was taken as a reference and the other was rotated by $\delta\theta = 3.9°$ and $\delta\phi = 6°$ with respect to the reference grain. The reciprocal space of the two grains is displayed in Figure 3, where the red lines correspond to the bulk crystal and the blue lines represent the rotated crystal.

**Algorithm 1** Grain simulation algorithm

```
procedure Reference Grain simulation
    function Define the parameters
        LC ← The lattice constant
        SF ← The structure factor
    function Generate possible HKL groups (HRange, KRange, LRange)
        for h in range (–HRange, HRange) do
            for k in range (–KRange, KRange) do
                for l in range (–LRange, LRange) do
                    CheckTheStructureFactor()
                    if True then
                        CaculateQxQyQz()
                    SaveAsReference()
        return DataArray[] ←The bulk datapoints
    function Generate The reciprocal space map
procedure Rotated Grains simulation
    repeat
    for i = 0: sizeof(DataArray[]) do
        ConvertToSpherical(RefArray[i])
        RotateToRef(RefArray[i], θ, ϕ)
        ConvertToCartesian(RefArray[i])
        SaveAsRotatedGrain()
    until number of required grains is reached
    return RotatedGrainsArray[][]
function Visualise the data
```

## *3.2 AT-map calculation*

To calculate the *AT-map*, each individual reflection in the experimental data is mapped based on its orientation stamp ($\epsilon_n$) with respect to its identical reflection in the simulated reference data. This is the 3D representation of the orientation stamp ($\epsilon_n$). It is important to distinguish between the two datasets: the simulated reference data and the experimental data. In the upcoming section, the prefix '*ref*' refers to the simulated dataset, while '*exp*' is used for the experimental data. The calculation of the AT-map can be performed using the following steps:

- Pairing each *exp* indexed reflection with its *ref* indexing-twin.
- Calculating the orientation stamp ($\epsilon$) for each pair.
- Analysing the three components $\alpha$, $\beta$, and $\gamma$ of the $\epsilon$ representing the projection of $\epsilon$ on the $x$–$y$, $y$–$z$, and $z$–$x$ planes.

**Figure 3** The simulated reciprocal map for bulk and rotated GaAs crystals with Miller indices range of −3 to 3 (see online version for colours)

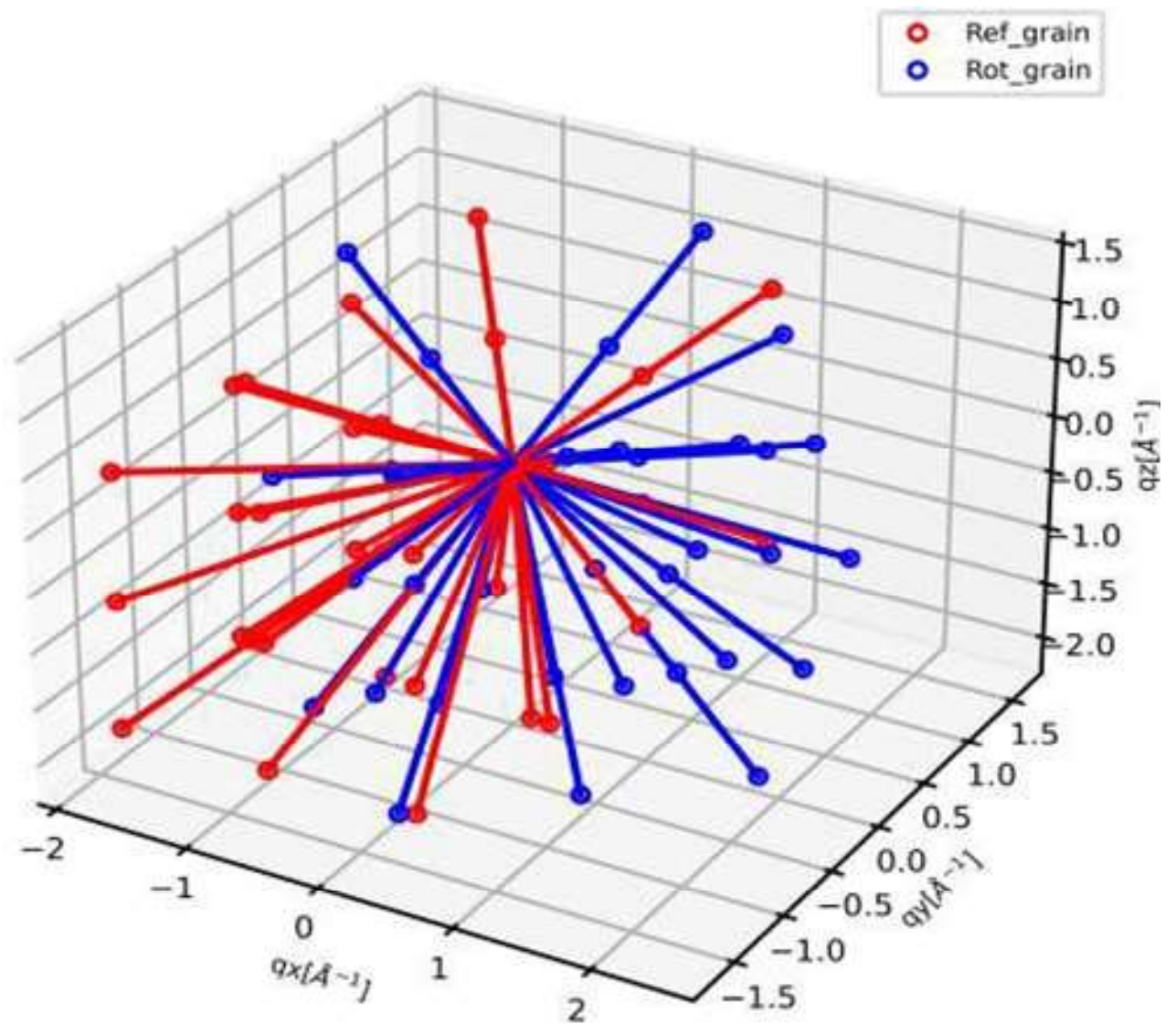

Notes: The red points represent the reciprocal lattice of the initial reference crystal grain while the blue points represent the reciprocal lattice of the rotated grain.

Upon processing the *AT-map*, the data is anticipated to be partitioned into multiple clusters of data points, where each cluster corresponds to reflections originating from a specific grain. Consequently, the number of clusters is equal to the number of grains probed in the experiment. This transformed data representation simplifies the problem into a conventional clustering scenario, enabling the application of ML techniques to address it effectively.

## 4 The twofold clustering-algorithm

Clustering tries to find structures in a dataset and group them in clusters by finding features that are similar between data points. In the context of this experiment, the following assumptions apply to describe the collected dataset:

a the possible orientations of grain are uniformly distributed on a sphere b all variables exhibit the same variance c the prior probability of finding ($k$), grains are uniform.

Given these assumptions and the statistical justification, various cluster methods, including DBSCAN (Khan et al., 2014; Deng, 2020), spectral clustering (Chen and Feng, 2012), K-means (Cui, 2020; Syakur et al., 2018), and hierarchical clustering (HC) (Murtagh and Contreras, 2012; Govender and Sivakumar, 2020), can be utilised. However, for our specific case at hand and the nature of our data, the K-means and HC clustering algorithms have been chosen. These selections are motivated by specific reasons that make them well-suited for the task:

K-means is a widely used and well-established clustering technique known for its simplicity in implementation and understanding. Its computational efficiency and ability to handle large datasets make it ideal for dealing with substantial amounts of data. Additionally, the straightforward interpretability of results, where each data point is assigned to a single cluster, is advantageous.

HC, on the other hand, allows for exploration of clusters at various levels of granularity. This adaptability is valuable when the number of clusters is uncertain and needs to be determined based on data characteristics.

Furthermore, the familiarity of our community with K-means and HC is a crucial factor in their selection. Researchers and practitioners' existing knowledge and experience with these methods facilitate easier adoption and comprehension of the clustering outcomes. Moreover, the widespread use of these algorithms in various applications further reinforces their effectiveness.

In summary, K-means and HC have been chosen due to their well-known attributes, computational efficiency, interpretability, flexibility, and widespread acceptance within the community. These factors make them highly suitable for the specific clustering task at hand.

To address the challenge of determining the unknown number of clusters (grains) using K-means, the elbow method has been proposed as a solution. This involves running K-means for a range of $k$ values and observing the curve of sum of squared errors ($SEE$) versus $k$. The 'elbow' point in the curve is considered an indication of the optimal number of clusters. However, using the elbow method to find the optimal number of clusters requires a closer look to make sure we understand how reliable it is for grouping grains in a polycrystalline energy dispersive Laue pattern. The elbow method is a common way to figure out the point where the improvement in grouping slows down, indicating the ideal number of clusters. While it is a widely used method, its reliability can depend on how the data is spread out and the inherent structure of the dataset. We need to be aware of potential issues, like situations where the elbow point is not clear or when the best number of clusters isn't well-defined because of complex data. Also, the elbow method assumes that clusters are round and of similar size, which might not match real-world datasets.

To enhance accuracy and performance, a hybrid approach combining K-means with the elbow method and HC has been proposed. The HC method determines the initial number of clusters ($k_{init}$), which is then validated by the elbow method. Subsequently, K-means is applied within a $k$ range starting from $k_{init} - 1$ until the elbow satisfies a specific threshold condition. This approach leverages the strengths of both techniques and is particularly effective for datasets with numerous clusters. The proposed model is presented in Algorithm 2 in the form of pseudo-code.

**Algorithm 2** Pseudo-code of the used twofold clustering-algorithm

**DataArray[]** ← The AT-map data
**function** Run the HC (**DataArray[]**)
    *return* $k_{init}$ ← The initial k value
*let* **IsElbow** = false
*let* $k^{+} = k_{init} - 1$
**while IsElbow** = false **do**
    **function** Run the K-mean ($k_{init}$)
        *return* $SEE_{init}$
    **function** Run the K-mean ( $k_{init}^{+}$ )
        *return* $SEE_{init}^{+}$
    **function** Check convergence ($SEE_{init}$, $SEE_{init}^{+}$ )
        *return* **IsElbow**
    **if IsElbow** == false **then**
        $k_{init} = k_{init}^{+}$
    **if IsElbow** == true **then**
        *return* **IsElbow** ← The final k value

## 5 Validation and discussion

Our methodology was evaluated on a diverse set of simulated datasets, encompassing four distinct scenarios:

1 Uncongested samples: These datasets consisted of only a few clusters and reflections (data points), resulting in a relatively sparse distribution.

2 Crowded samples: These datasets contained a larger number of data points and clusters, leading to a more densely populated distribution.

3 Non-homogeneous crowded samples: In these datasets, the number of clusters and data points was extremely high, with a non-uniform distribution across the clusters, creating a challenging scenario.

4 Constrained reciprocal-space samples: These datasets represented a specific slice through reciprocal space as detected by the instrument (i.e., the detector), introducing constraints in the data distribution.

The simulated datasets for each scenario were generated using the simulation approach outlined in Algorithm 1. Below, we provide detailed information about the simulated datasets in each scenario.

### 5.1 Uncongested samples

As an example for the uncongested samples, a dataset of three GaAs grains was simulated with a HKL range of 3:3. Each grain has a different orientation with respect to the reference grain. The orientation angles were randomly sampled following a normal distribution. Figure 4(a) shows the 3D reciprocal mapping while Figure 4(b) displays the corresponding *AT-map* representation (after data preparation steps). This AT-map representation has been fed to our twofold classifiers giving the following results:

- The HC: We employ a threshold of 3° per cluster representing the angular resolution of the detector. The HC votes for $k = 3$ to be the optimal clusters number and the dendrogram of the decision tree is shown in Figure 4(c).
- K-mean with the elbow method: Starting with a value of $k_{init} = 2$, the algorithm voted for $k = 3$ to be the elbow point which agrees with the result of the HC classifier. Figure 4(d) displays the SSE for different values of $k$ within the examined range.

The resulting clustered AT-map is shown in Figure 4(e). The data-points are grouped in three clusters with the defined centroids and a list of labelled reflections is generated showing each reflection with its corresponding grain (cluster) identifier.

### 5.2 Crowded samples

Another validation use-case is shown using a dataset of 20 GaAs grains which are randomly oriented with respect to the reference grain and with HKL range of 3:3. The visualisation of the reciprocal space data points is shown in Figure 5(a). Following the previous approach, the *AT-map* was given to the HC clustering algorithm which voted for $k = 20$. Then, the supervisor with a $k_{init} = 19$ approved the result. As an output, the clustered AT-map was generated showing the 20 clusters [Figure 5(b)].

### 5.3 Non-homogeneous crowded samples

A dataset of randomly oriented 50 GaAs grains with HKL range of 20:20 was generated. Each grain contains a random number of reflections between 1 to 6. A single reflection was generated with random HKL values. The reciprocal representation is shown in Figure 5(c). The voted and approved result from our clustering approach is, as expected, 50 grains. The clustered AT-map is shown in Figure 5(d).

### 5.4 Non-homogeneous crowded samples with constrained reciprocal space

To generate non-homogeneous crowded samples with constrained reciprocal space, the previous GaAs dataset of 50 grains was used but a constrained reciprocal space that is defined by the detector active area. This constraint ensured that the reflections were limited to the region captured by the detector during the experiment. The reciprocal space of the filtered data points is shown in Figure 5(e). Both the propose and the supervisor voted for 47 clusters (grains). The clustered AT-map is shown in Figure 5(f).

# 6 Evaluation

## *6.1 Accuracy*

Prior to deploying our twofold approach with real experimental data, we conducted an evaluation and testing phase to assess its limitations and accuracy. To measure the clustering accuracy, we used the adjusted rand index (ARI) as a metric, which considers both true positives and true negatives and is adjusted for chance.

**Figure 4** Data points (reflections) generated by three grains, (a) shows the representation of the reciprocal space (b) is the *AT-map* representation, while (e) is the clustered one (c) (d) are the outcomes of the clustering system (see online version for colours)

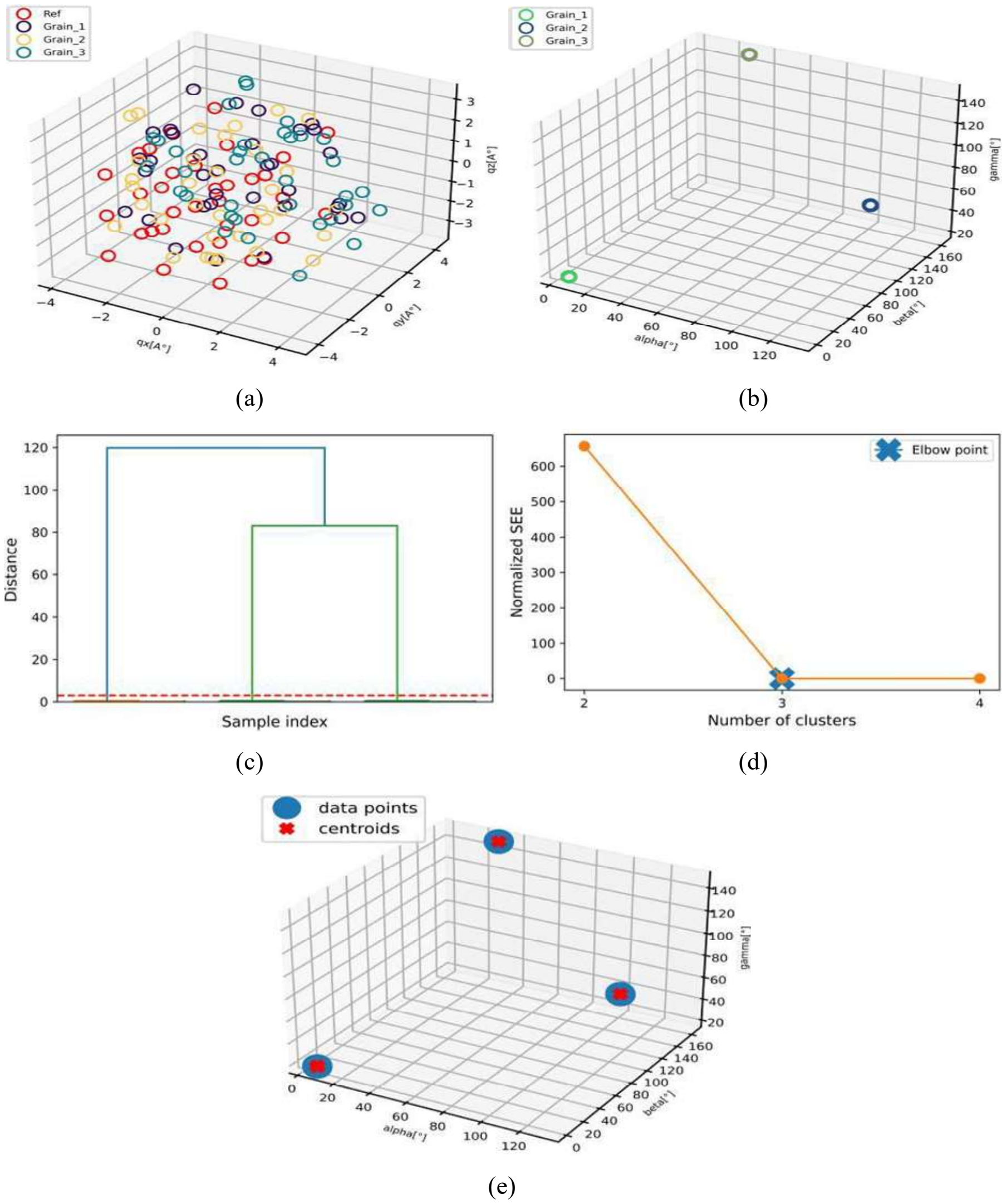

**Figure 5** The representations of the reciprocal space and clustered *AT-map* for 20 grains are shown in data points (reflections) (a) (b) respectively; similarly, data points (reflections) (c) (d) depict the reciprocal space and clustered *AT-map* for 50 grains; additionally, (e) (f) illustrate the constrained reciprocal space and clustered *AT-map* (see online version for colours)

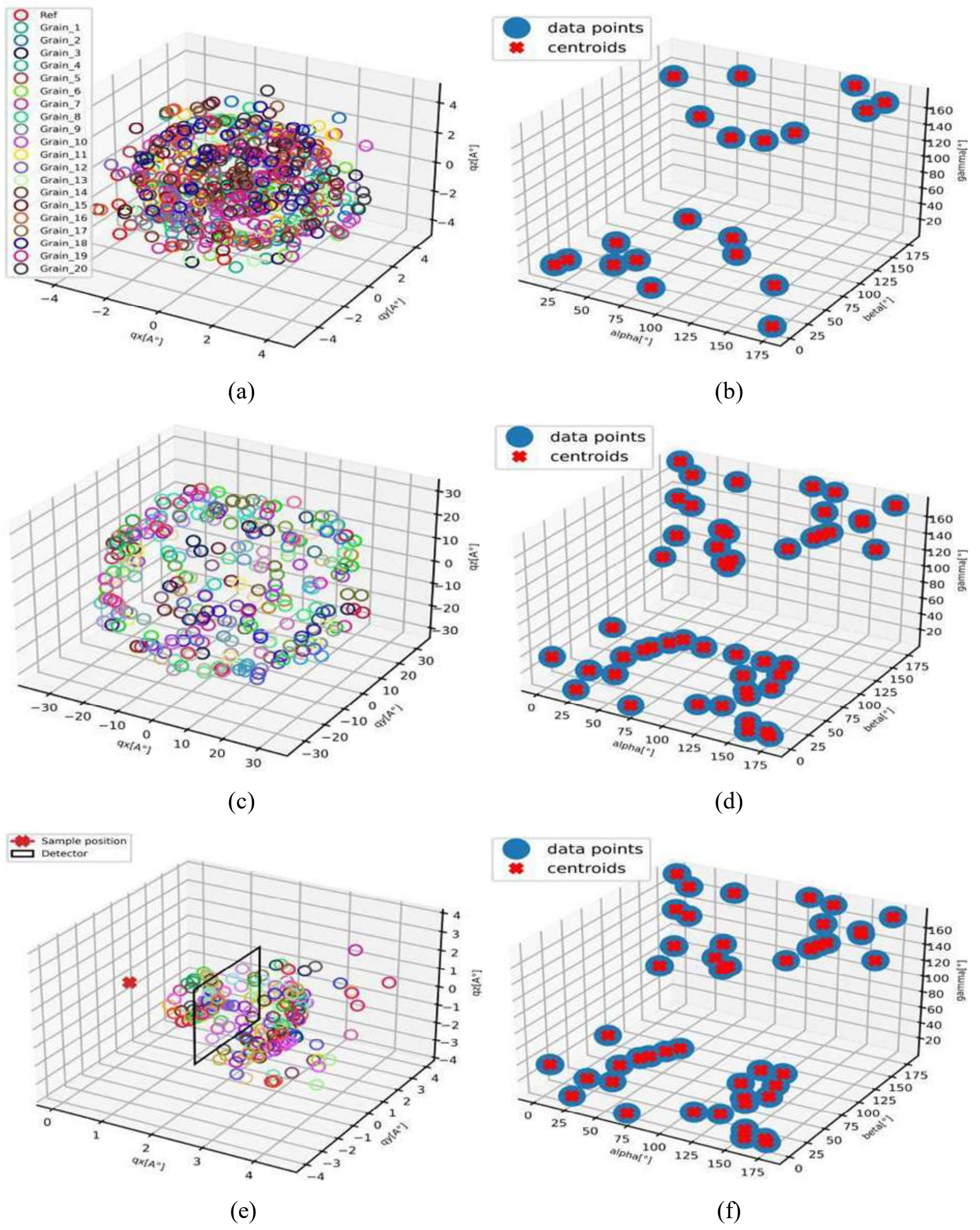

The testing process was carried out in three steps:

1 In this case, we are examining situations where the orientation stamp between grains is greater than 3°, and the number of reflections per grain falls between 1 and 6. As illustrated in Figure 6, the ARI achieves its highest value (100%) for a number of grains below 200. However, as the number of grains represented within the sample surpasses 400, the ARI begins to decline, reaching its lowest value.

**Figure 6** The ARI of the twofold approach vs. number of grains with orientation stamp of grains > 3°, and 1 < the number of reflections per grain > 6 (see online version for colours)

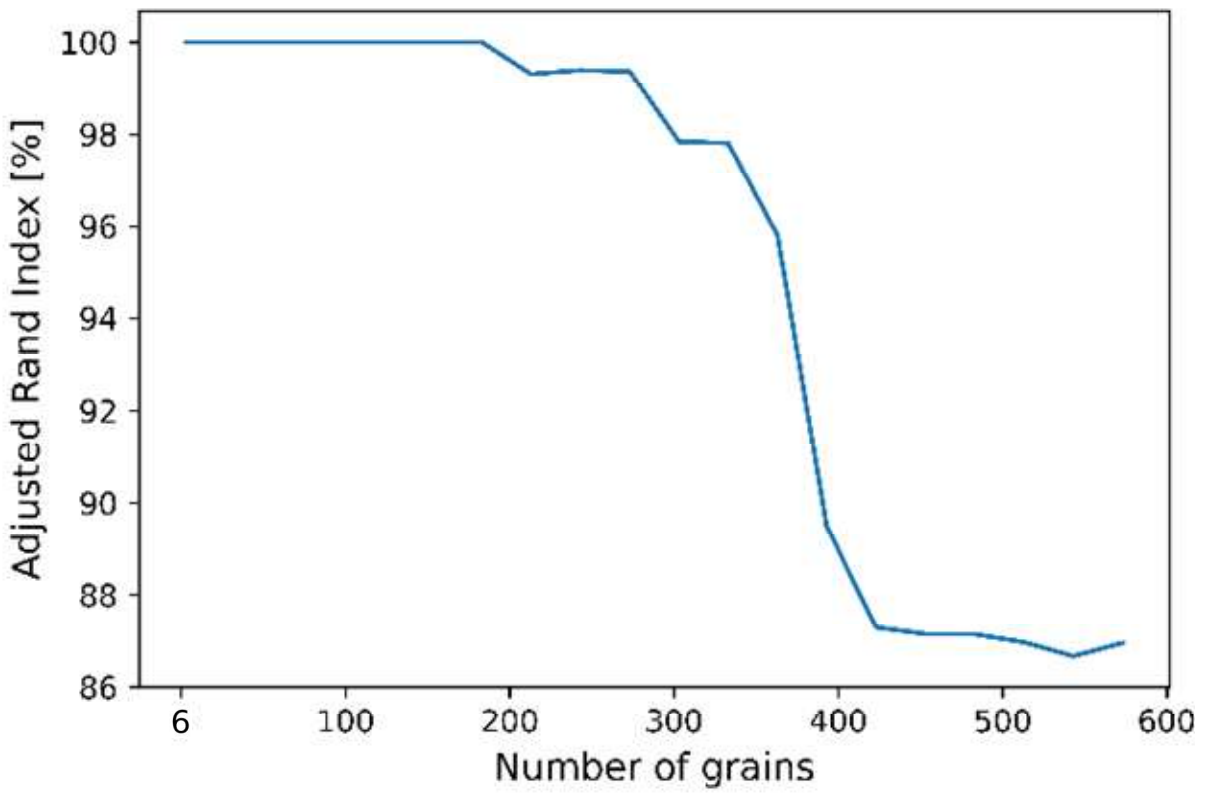

2 In this scenario, we are exploring situations where the orientation stamp is greater than 3°, and the number of reflections per grain falls between 20 and 10. The objective of increasing the number of reflections per grain is to enhance the data's density, thereby increasing the overall crowding of the dataset. According to Figure 7, the clustering system achieves an ARI of over 98% for datasets containing less than 100 grains. However, the ARI drops significantly for samples containing more than 150 grains.

**Figure 7** The ARI of the twofold approach vs. number of grains with orientation stamp of grains > 3°, and the number of reflections per grain < 10 (see online version for colours)

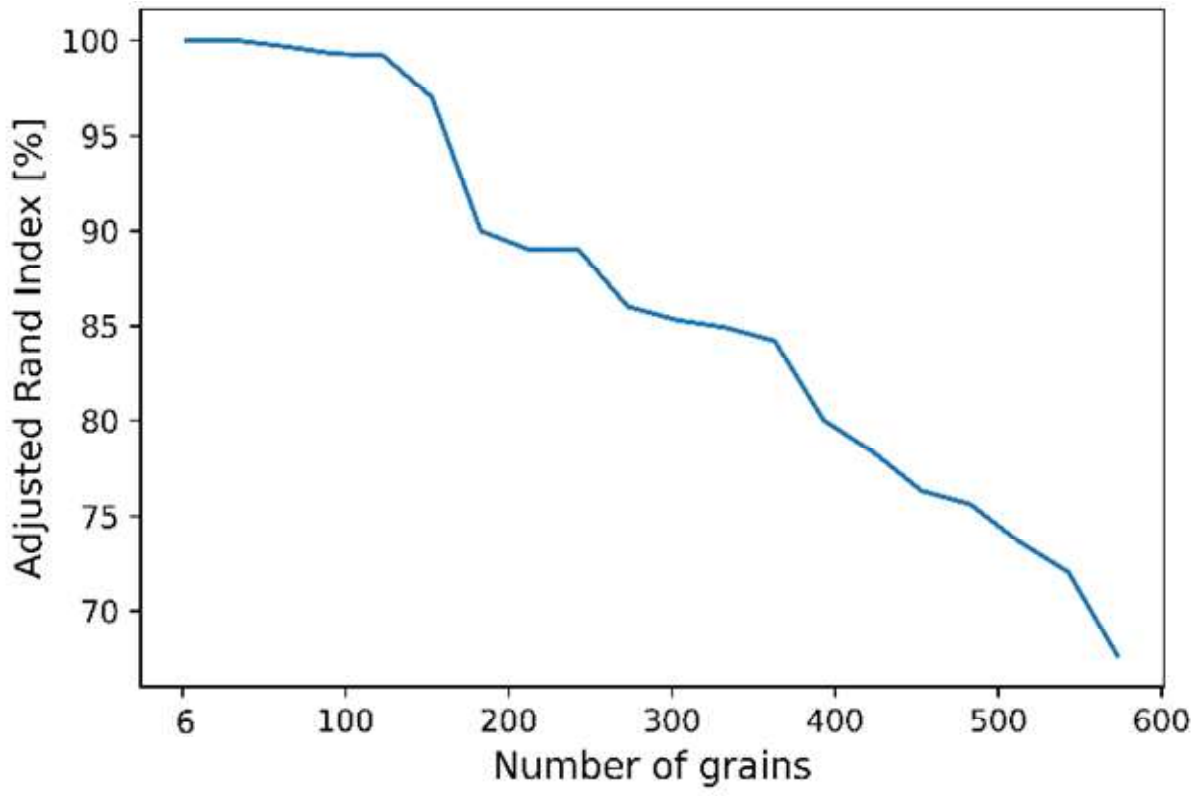

3 In this case, we are examining instances where the orientation stamp of grains is less than 3°, and the number of reflections per grain is fixed at 6. As depicted in Figure 8, orientation stamps of 2° and 3° exhibit high accuracy for samples with grains less than 200. However, the orientation stamp of 1° yields significantly lower accuracy, irrespective of the number of grains within the sample.

**Figure 8** The ARI for various orientation stamps (1°, 2°, and 3°) (see online version for colours)

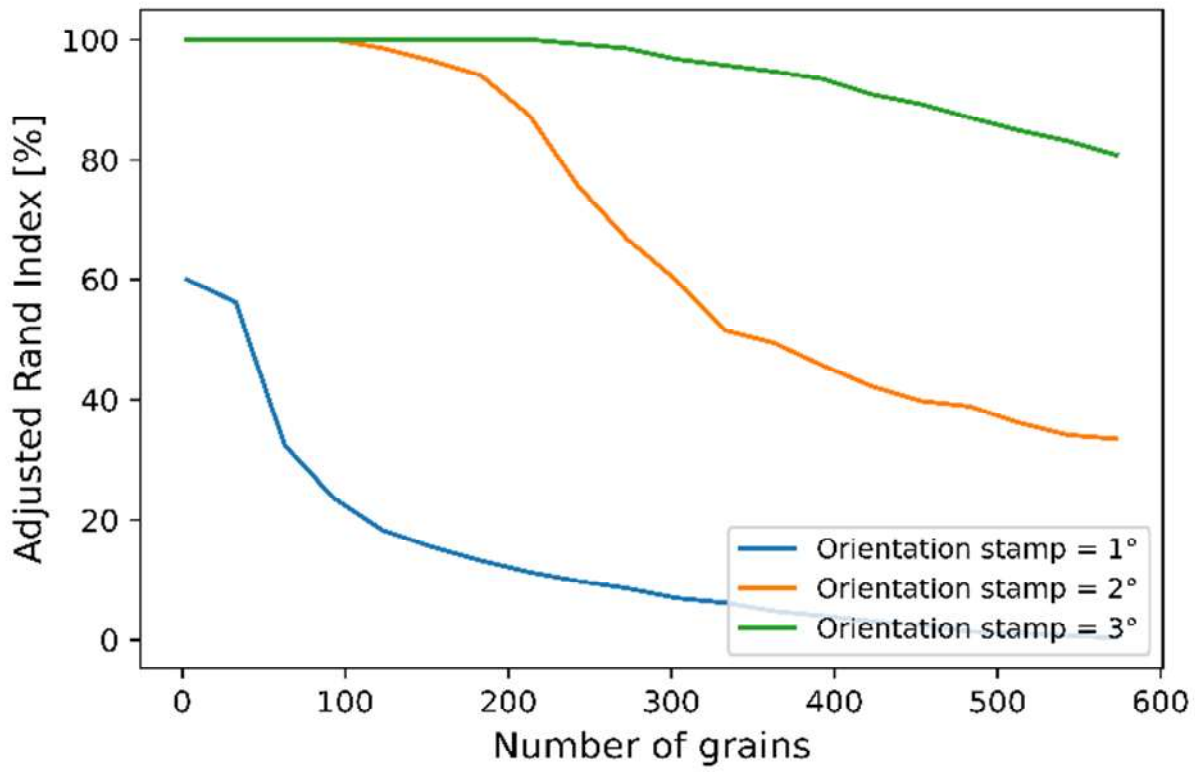

The conducted tests have yielded compelling evidence supporting the reliability and effectiveness of this twofold approach. Notably, the approach demonstrates high accuracy when dealing with samples containing fewer than 200 grains, and it successfully handles orientation stamp differences of at least 2°. The results indicate that the method performs optimally within these specified conditions. Moreover, it is crucial to acknowledge that the presented limitations, particularly the constraints imposed by the spectral resolution of the utilised detectors, are acceptable trade-offs in practical scenarios. Spectral resolution limitations can pose challenges in accurately resolving fine differences in orientations, especially in samples with a higher number of grains or narrower orientation differences. Given these constraints, the twofold approach still provides valuable and reliable insights, making it well-suited for applications involving samples with moderate grain counts and orientation differences. It is essential to be aware of these limitations when interpreting the results, as they can guide researchers in selecting appropriate sample sizes and orientation ranges for optimal accuracy and performance.

## 6.2 *Performance*

The performance evaluation of the algorithm involved a comprehensive latency analysis of its three primary components: grain simulation, AT-map calculator, and classification. This analysis aimed to assess the efficiency and effectiveness of each component in handling various aspects of the data processing pipeline.

To begin with, the grain simulation component was subjected to testing to measure the elapsed time required for generating reference grains with varying HKL ranges. Figure 9 demonstrates the elapsed time by the grain simulator to generate the reference grain with different HKL ranges. Notably, the practical EDLD experiments typically have HKL ranges that do not exceed 20:20 due to inherent geometric limitations in experimental setups, such as beam energy, detector size, and detector quantum efficiency. The results demonstrated that the maximum latency for simulating a reference grain remained within a negligible time, usually less than 1 second. This rapid grain simulation process ensures the algorithm's responsiveness and makes it suitable for real-time applications.

**Figure 9** The elapsed time to simulate the reference grain vs. the HKL ranges (see online version for colours)

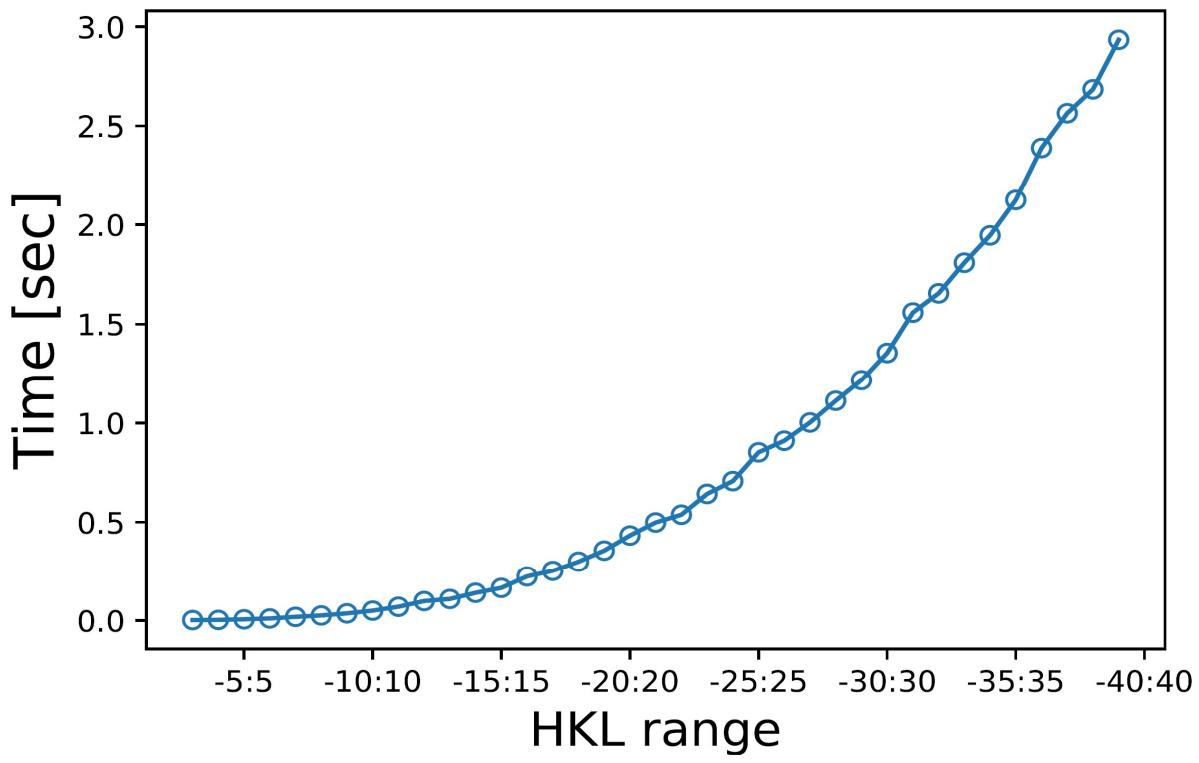

Next, the AT-map calculator, a crucial part of the algorithm responsible for generating *AT-maps* from diffraction data, underwent thorough investigation. The analysis was conducted using a reference grain with an HKL range of 20:20, representing typical experimental scenarios. It was observed that the performance of the *AT-map* calculator exhibited sensitivity to the density of Laue patterns in the input data, as it is shown in Figure 10. In cases where the Laue patterns were highly crowded, containing over 300 reflections, the AT-map calculator experienced a slight decrease in speed. However, it is important to note that this performance issue was less pronounced when processing datasets with fewer than 300 reflections.

**Figure 10** The latency of the At-map calculator (see online version for colours)

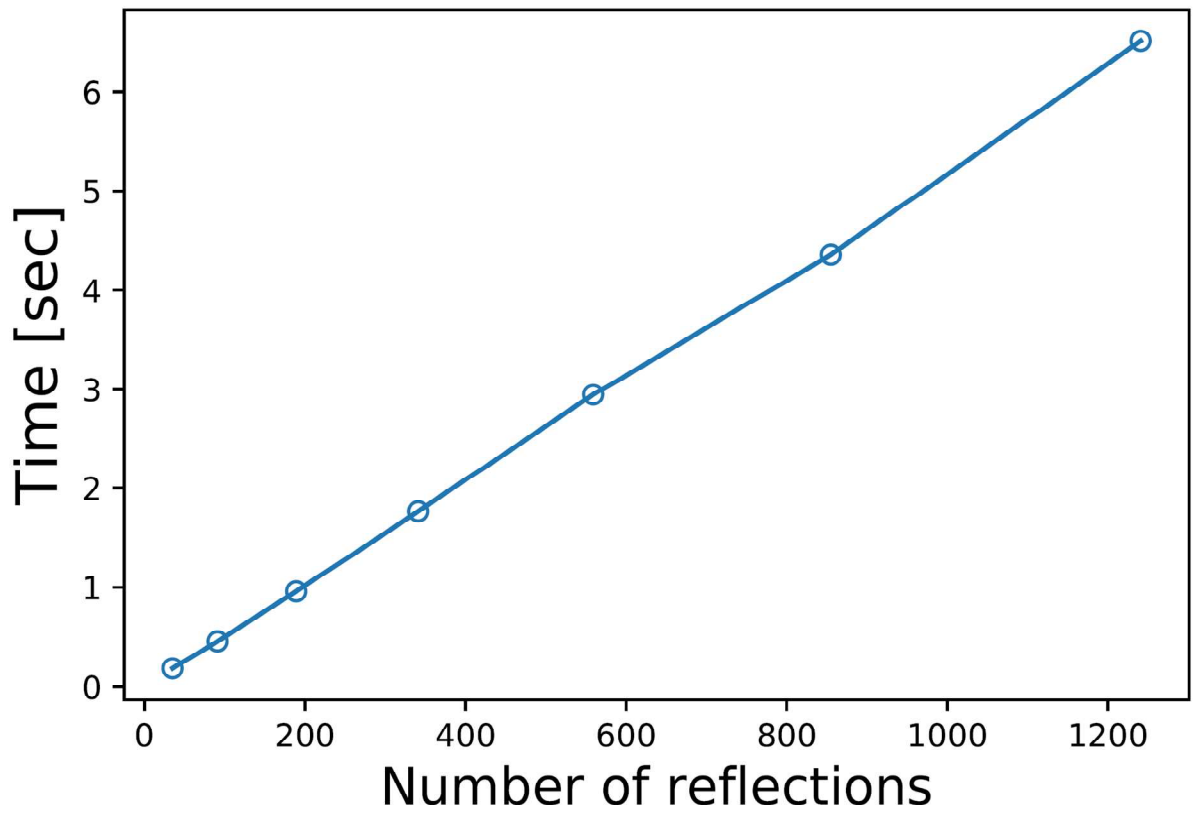

Lastly, the clustering twofold system was also assessed. The results, as it is demonstrated in Figure 11, indicated that the clustering procedure had minimal impact on overall latency, contributing negligibly to the computational overhead.

**Figure 11** Assessment of results from experimental data, (a) shows the indexed Laue pattern of GaAs single crystal, where (b) is the clustered *AT-map* of this dataset; the red x represents the centroid of the predicted cluster; the indexed Laue pattern of Ni polycrystalline sample is in (c) and the clustered *AT-map* of the Ni dataset is in (d) (see online version for colours)

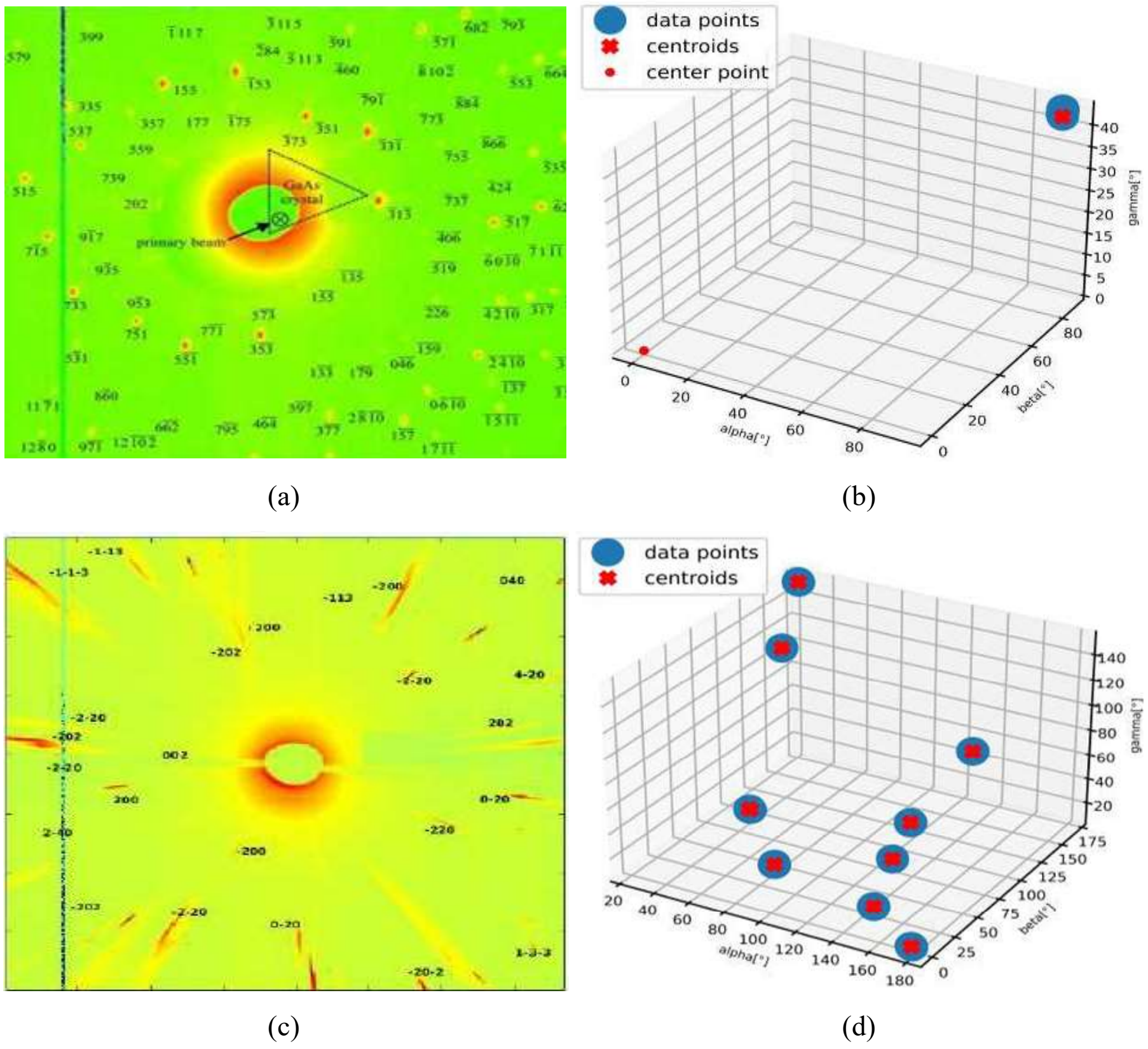

## 7 Evaluating on experimental data

After successfully validating the algorithm's performance with simulated data, we advanced to the next phase where we applied the algorithm to datasets obtained from real experimental measurements. In this phase, the resulting clustering outcomes and corresponding *AT-maps* were meticulously analysed and compared against previous analysis results. This comparison aimed to evaluate the algorithm's effectiveness and its ability to accurately capture patterns and structures in the real-world data. In this context, we utilised two distinct experimental datasets for our analysis.

### *7.1 Single crystal material*

A dataset was obtained from a single crystal sample of GaAs. The experimental procedure involved mounting the single crystalline GaAs sample at the EDDI beamline of the BESSY II storage ring in Berlin. The corresponding Laue pattern was captured using a pnCCD in transmission geometry (Send et al., 2016). In Figure 11(a), you can observe the Laue pattern recorded in a single exposure, with a sample detector distance of 41 mm (STD = 41 mm). The Miller indices range was limited to between –10 and 10.

Before clustering, the data was previously analysed and confirmed to represent a single grain. The reciprocal space data was then subjected to our twofold clustering system. First, the calculated AT-map was generated and utilised as input for the clustering algorithm. Employing the HC technique, all data points were merged into a single cluster, resulting in $k = 1$. Subsequently, the elbow k-means model with $k_{init} = 1$ also suggested a single cluster. Consequently, this dataset was predicted to contain only one cluster, as depicted by the centroid shown in Figure 11(b). The agreement between the clustering results and the prior analysis confirms the presence of a single grain in the examined Laue pattern.

### *7.2 Polycrystalline material*

In this investigation, we obtained a dataset from a polycrystalline sample by performing an EDLD experiment using a polycrystalline nickel wire at the EDDI beamline of BESSY II (Shokr, 2019). The pnCCD camera was used to capture the Laue pattern, and after carefully selecting and processing several reflections, we visualised the Laue pattern along with the corresponding assigned reflections in Figure 11(c).

In a previous analysis (Shokr, 2019), it was determined that this reflection pattern was generated by nine different grains. To validate and corroborate these findings, we subjected the dataset to the same analysis procedure as described earlier. The 3D AT-map was prepared and utilised as input for the clustering algorithm. By examining the threshold line, we identified nine intersection points with the plotting, indicating that $k_{init} = 9$. Additionally, the elbow k-means analysis also suggested an elbow point at 9. Consequently, the clustering algorithm successfully identified nine distinct grains in this dataset, thus reaffirming the results of the previous analysis. The clustered AT-map, displaying the distribution of the identified grains, can be observed in Figure 11(d).

These findings demonstrate the robustness and accuracy of our clustering approach in successfully identifying and distinguishing individual grains in complex polycrystalline datasets. The clustering results provide valuable insights into the microstructure and crystallographic characteristics of the sample under investigation. The ability to analyse and classify grains in single crystal and polycrystalline materials holds significant implications for material science, materials engineering, and related fields.

## 8 Conclusions and future work

This paper presents a ML-based solution for identifying two grain-related clustering of Laue spots in both single crystalline and polycrystalline energy dispersive Laue patterns. The study highlights the significance of data point representation in achieving successful clustering outcomes. Specifically, the adoption of the *AT-map* representation is proposed as an innovative approach for preparing EDLD datasets, which proves to be effective in accurately representing the underlying grain structures.

Moreover, to enhance the performance and accuracy of the clustering procedure, a combination of two classifiers, namely HC and elbow k-means, is employed. This hybrid approach exhibits improved clustering results, leading to a more reliable identification of grain-specific reflections.

The ARI of the clustering system was rigorously tested in various scenarios, assessing its performance under different conditions. Through extensive testing, the strengths and limitations of this clustering system were thoroughly evaluated.

In scenario-specific analyses, the ARI was calculated to quantify the agreement between the clustering results and the ground truth. This allowed for a comprehensive assessment of the clustering system's accuracy and effectiveness in capturing underlying patterns and structures in the data.

An essential aspect addressed in the paper is the latency analysis, which emphasises the real-time properties of the developed algorithm. The efficient processing time of the proposed algorithm showcases its potential for practical applications, where quick and accurate grain identification is crucial.

The experimental findings and insights presented in this paper contribute significantly to the field of materials science and electron diffraction analysis. The demonstrated ML-based approach opens up new possibilities for non-destructive and automated grain identification, which is vital in various scientific and engineering applications. Additionally, the innovative AT-map representation introduces a novel perspective for data representation in EDLD, potentially impacting future research and development in this domain. Overall, the comprehensive analysis and positive results of the proposed method establish its relevance and potential for advancing grain-selective Laue reflection analysis in diverse material systems.

As part of future work, our focus will be on exploring and evaluating new clustering algorithms in the context of our research. We recognise the importance of continuously seeking advancements and innovations in clustering techniques to enhance the performance and capabilities of our system.

To achieve this, we will conduct a comprehensive benchmarking study, where we will compare the performance of various state-of-the-art clustering algorithms, such as DBSCAN or spectral clustering with the system we have introduced. Examining these alternatives will shed light on whether different techniques yield comparable or superior outcomes in terms of accuracy, efficiency, and robustness.